# THE STARK INTERACTION OF IDENTICAL PARTICLES WITH VACUUM ELECTROMAGNETIC FIELD AS QUANTUM POISSON PROCESS SUPPRESSING COLLECTIVE SPONTANEOUS EMISSION

## A.M.Basharov

*RRC «Kurchatov Institute», Moscow, Russian Federation,* e-mail: basharov@gmail.com

The effective Hamiltonian describing resonant interaction of an ensemble of identical quantum particles with a photon-free vacuum electromagnetic field has been obtained with allowance for the second-order terms over the coupling constant (the Stark interaction) by means of the perturbation theory on the basis of the unitary transformation of the system quantum state. It has been shown that in the Markov approximation the effective Hamiltonian terms of the first-order coupling constant are represented as the quantum Wiener process, whereas the second-order terms are expressed by the quantum Poisson process. In the course of investigation it was established that the Stark interaction played a significant role in the ensemble dynamics, thus influencing the collective spontaneous decay of the ensemble of an appreciably high number of identical particles. New fundamental effects have been discovered, i.e., the excitation conservation in a sufficiently dense ensemble of identical particles and superradiance suppression in the collective decaying process of an excited ensemble with the determined number of particles.
PACS numbers 42.50.Ct, 42.50.Hz, 42.50.Lc, 42.50.Nn, 02.50.Fz, 02.50.Ga

## 1. INTRODUCTION

As far as the dynamics of excited atoms in a resonant broadband electromagnetic field is concerned, it reduces to the below stated notions. Atom interaction with a photon-free vacuum electromagnetic field will both give rise to the spontaneous transition from the excited to the ground state, with photon being emitted, and the Lamb shift of atomic levels [1]. The spontaneous decay of an excited state is effectively described by a two-level quantum particle model [2]. An ensemble of identical excited two-level quantum particles localized in a small volume radiates a coherent electromagnetic pulse with the time delay. The pulse intensity is proportional to the square of the particle number and the pulse duration is inversely proportional to the particle number. As the number of particles in an ensemble increases, the pulse duration and its time delay decrease, with intensity enhancing. Nowadays, this phenomenon discovered by Dicke [3] is known as superradiance and subject to comprehensive investigations [2-4].

In conventional conditions the Lamb shift of resonant atomic frequency in a vacuum electromagnetic field is negligibly small in comparison with the characteristic Rabi frequency describing radiation transitions between quantum resonant levels. Therefore, in the dynamics of spontaneous decay the Lamb shift is taken into account only in the renormalization of resonant transition frequency and does not influence the decay in any way. The Lamb shift infinitesimality indicated is due to the fact that the Rabi frequency is of the first order value of the coupling constant with an electromagnetic field, whereas the Lamb shift is of the second order of this constant.

Apart from the Lamb shift, the second order of the coupling constant with an electromagnetic field is also characteristic of the Stark interaction of an atom with a vacuum electromagnetic field. In a classical field, the Stark interaction is displayed in the form of high-frequency Stark level shifts, making a significant impact on non-linear optics phenomena [5].

For a vacuum electromagnetic field, the average value of the Stark interaction operator describing the quantum Stark level shifts is equal to zero. Nevertheless, at sufficient intensity the Stark interaction influences the spontaneous decay of a single quantum particle, producing the suppression of spontaneous emission and an additional energy shift of a decaying level [6]. For the Stark interaction value of a single quantum particle to be sufficiently effective, it is necessary to follow non-conventional conditions, e.g. the two-quantum resonance, with only one photon being emitted [6]. Then the Stark interaction is of the same order as that of the two-photon Rabi frequency.

In the present paper it has been shown that the Stark interaction becomes extremely essential in an ensemble of a sufficiently large number of identical atoms in spite of the fact that the value of the same Stark interaction is small for a single atom. That is why the Stark interaction may prove to influence the collective spontaneous decay of an ensemble of identical quantum particles. In contrast to Dicke superradiance [2-4] there have been discovered some cases of superradiation in which the increasing of the particle number yields the increasing of pulse duration and pulse delay due to the Stark interaction, whereas the rate of collective spontaneous emission reduces. Moreover, it has been found that there exists the "critical" atom number in an ensemble, the superradiance of which is fully suppressed. If the atom number of an ensemble exceeds this "critical" atom number, the collective spontaneous emission, once started, brings to a halt, i.e. the atomic ensemble gets stabilized in an excited state. These effects are due to the suppression of spontaneous emission of a single particle by the Stark effect described in Ref. [6]. As opposed to Ref. [6] a new parameter defining the critical importance of the Stark interaction is the particle number in a decaying ensemble. Besides, the new results indicated above are related to collective spontaneous emission from an ensemble of any number of identical quantum particles if the Stark interaction of a single quantum particle with a vacuum electromagnetic field turns out to be sufficiently high for a number of particular reasons.

The results shown in the paper have been obtained on the basis of the derived effective Hamiltonian with the related non-Langevin quantum stochastic differential equation (QSDE). The Stark interaction is shown to be presented by the quantum Poisson process which is responsible for the non-Langevin type of the QSDE. Therefore, the superradiance allowing for the Stark interaction with electromagnetic vacuum is to be referred to as non-Langevin supperradiance in contrast to the Dicke superradiance [2-4], which is of Langevin type.

Derivation of master equations can be made by various methods (see Refs.[7-9]), whereas the QSDE method [5,6,8,11-14] is not only most elegant and straightforward but also represents part and parcel of mathematically correct description of open systems of a particular class. (The spontaneously emitting excited atoms give an important example of open systems). The paper emphasizes the important role played by the effective Hamiltonian picture of open system for consistent analysis of the system dynamics. Conventionally, the derivation of the master equation by the QSDE method was based on the Hamiltonian in the rotating frame approximation [12,13]. Such a Hamiltonian presented the initial electro-dipole interaction Hamiltonian without fast oscillating terms in the Dirac picture. The indirect application of the QSDE method to the initial electro-dipole interaction Hamiltonian gives rise to an unexpected and contradictory observational result [15]. The relaxation went missing in that case as if the two-level excited atom had not undergone any radiative decay. Thus, in the QSDE method, the problem of an effective Hamiltonian arises. The basic assumptions regarding the interaction of an open system with its environment must be applied not to any Hamiltonian (including any general and exact Hamiltonian) but to an effective Hamiltonian. The systematic principle of the effective Hamiltonian derivation as well as its applicability has been formulated in the present paper. Such an approach allows making a straightforward derivation of not only the rotating frame approximation Hamiltonian but to that of basic term describing the Stark interaction of quantum particles with a broadband quantized electromagnetic field responsible for the non-Langevin type of superradiance. The same very approach imposes restrictions on the further application of the master equation in the process of investigating of open system dynamics which were neglected in several previous works, leading to obtaining incorrect results.

Our consideration of the Stark interaction can prove to be useful for solving other similar problems in which it is necessary to take into account the operator represented by the quantum Poisson process.

The paper is organized as follows. In section 2 the unitary Hamiltonian transformation together with the perturbation theory is used to derive the effective Hamiltonian and to introduce the effective Hamiltonian picture. In section 3 the QSDE notion is involved for the Markov approximation, and the non-Langevin evolution operator is derived for a photon-free quantized electromagnetic field. Section 4 is devoted to the derivation of the master equation and its representation in the Lindblad form. The fifth section considers the spontaneous decay of singly excited ensemble symmetrical over particle permutation. The sixth section is concerned with the peculiarities of the non-Langevin collective spontaneous decay of the fully excited ensemble and the introduction of critical values at which the atomic ensemble emission is fully suppressed. The final section considers the effective Hamiltonian picture and QSDE as the basis of systematic investigation of open systems including physical systems containing fast and slow subsystems.

## 2. EFFECTIVE HAMILTONIAN PICTURE OF THE PROBLEM

Let us consider an ensemble of $N_a$ identical motionless atoms interacting with quantized broadband electromagnetic field. The atoms are localized near a point $\vec{r}=0$ in a small volume the size of which is much smaller than the characteristic wavelength of an electromagnetic field. An initial Hamiltonian of such a system in an electro-dipole approximation

$$H^{Ini} = H^A + H^F + H^{Int} \tag{1}$$

consists of the sum of Hamiltonian $H^A$ of isolated atoms, electromagnetic field Hamiltonian $H^F$ and interaction operator of atoms with an electromagnetic field $H^{Int}$,

$$H^A = \sum_{i,j} E_j |E_j>^{(i)} <E_j|^{(i)}, \quad H^F = \sum_{\vec{q}} \hbar\omega_{\vec{q}} b_{\vec{q}}^+ b_{\vec{q}}, \quad H^{Int} = \sum_{\vec{q}} \Gamma_{\vec{q}} (b_{\vec{q}}^+ + b_{\vec{q}}) \sum_{i,kj} d_{kj} |E_k>^{(i)} <E_j|^{(i)}, \tag{2}$$

$$\sum_j |E_j>^{(i)} <E_j|^{(i)} = 1^{(i)}, \quad <E_j|^{(i)} E_k>^{(i)} = \delta_{jk},$$

where $|E_j>$ is an atomic quantum nondegenerate state of energy $E_j$, $d_{kj} = <E_k|d|E_j>$ is the matrix element of the atomic dipole moment operator $d = \sum_{kj} d_{kj} |E_k><E_j|$. Atomic states are characterized by a definite parity, so that $<E_k|d|E_k>=0$. The upper index of state vectors designates the state space of an *i*th atom; the sum is over all ensemble atoms. Annihilation and creation operators of photon with wave vector $\vec{q}$ and frequency $\omega_{\vec{q}}$ are given by $b_{\vec{q}}$ and $b_{\vec{q}}^+$, $[b_{\vec{q}}, b_{\vec{q}'}^+] = \delta_{\vec{q}\vec{q}'}$, $\omega_{\vec{q}} = qc$. The atomic coupling with a conventional three-dimensional electromagnetic field is characterized by the value $\Gamma_{\vec{q}} = (2\pi\hbar qc/\ell^3)^{1/2}$, with $\ell^3$ as quantization volume. The recoil effects and polarization photon states are neglected. The dipole-dipole interaction of identical atoms is neglected analogous to conventional theory of superradiance [2-4]. The justification of electro-dipole approximation for a single two-level atom can be found in Ref. [1].

The state vector $|\Psi(t)>$ of the system containing atoms and quantized electromagnetic field in the interaction (Dirac) picture obeys the Schrödinger equation

$$i\hbar \frac{d}{dt} |\Psi(t)> = H^{Int}(t) |\Psi(t)>, \tag{3}$$

$$|\Psi(t)> = \exp(i(H^A + H^F)t/\hbar) |\Psi>,$$

$$H^{Int}(t) = \exp(i(H^A + H^F)t/\hbar) H^{Int} \exp(-i(H^A + H^F)t/\hbar) =$$

$$= \sum_{\vec{q}} \Gamma_{\omega_{\vec{q}}} (b_{\vec{q}}^+ e^{i\omega_{\vec{q}}t} + b_{\vec{q}} e^{-i\omega_{\vec{q}}t}) \sum_{i,kj} d_{kj} |E_k>^{(i)} <E_j|^{(i)} e^{i\omega_{kj}t}, \quad \omega_{kj} = (E_k - E_j)/\hbar,$$

where $|\Psi>$ is the system state vector in the Schrödinger picture.

According to unitary symmetry of quantum theory let us make a unitary transformation,

$$|\tilde{\Psi}(t)> = U(t)|\Psi(t)>. \tag{4}$$

Transition from vector $|\Psi(t)>$ towards vector (4) is accompanied with the Hamiltonian change

$$\tilde{H}^{Int}(t) \equiv \Im(H^{Int}(t)) = U(t)H^{Int}(t)U(t)^+ - i\hbar U(t)\tfrac{d}{dt}U(t)^+. \qquad (5)$$

Here, the description of quantum system is expressed with the help of the Schrödinger equation with a transformed Hamiltonian (5):

$$i\hbar \frac{d}{dt}|\tilde{\Psi}(t)> = \tilde{H}^{Int}(t)|\tilde{\Psi}(t)>. \qquad (6)$$

Consider a unitary operator $U(t)$ in terms of the Hermitian operator

$$U(t) = e^{-iS(t)}, \quad S(t)^+ = S(t), \qquad (7)$$

in order to use the Baker-Hausdorff formula for an arbitrary operator $O$

$$e^{-iS}Oe^{iS} = O + \frac{(-i)}{1!}[S,O] + \frac{(-i)^2}{2!}[S,[S,O]] + \frac{(-i)^3}{3!}[S,[S,[S,O]]] + \dots.$$

A transformed Hamiltonian (5) and operator $S(t)$ are expanded in series of the coupling constant

$$S(t) = S^{(1)}(t) + S^{(2)}(t) + \dots, \quad \tilde{H}^{Int}(t) = \tilde{H}^{Int(1)}(t) + \tilde{H}^{Int(2)}(t) + \dots, \qquad (8)$$

where the upper index signifies the expansion order of the coupling constant. Substituting (7), (8) into (5) with account of the Baker-Hausdorff formula and equating the expression of the same order, we have

$$\tilde{H}^{Int(1)}(t) = H^{Int}(t) + \hbar dS^{(1)}(t)/dt, \qquad (9)$$

$$\tilde{H}^{Int(2)}(t) = -\frac{i}{2}[S^{(1)}(t), H^{Int}(t)] - \frac{i}{2}[S^{(1)}(t), \tilde{H}^{Int(1)}(t)] + \hbar \frac{dS^{(2)}(t)}{dt}, \qquad (10)$$

$$\dots$$

Expansion (8) and formulae (9)-(10) define unitary transformation (4)-(8) in a unique way by requiring the absence of fast time varying factors in relevant terms in the interaction picture. We characterize this transformation as the transition to an effective Hamiltonian picture. The latter similar to Heisenberg and Dirac (interaction) pictures is closed because the repetitive (or *n*th fold) unitary transformations $\Im$ leave an effective Hamiltonian «fixed», $\tilde{H}^{Int}(t) = \Im\{\tilde{H}^{Int}(t)\}$, since it is a fixed point of sequential identical unitary transformations $\Im$.

Note that the unitary transition to an effective Hamiltonian picture can be made in the Schrödinger picture as well. Then the basic transformation formulae will be the following

$$|\tilde{\Psi}> = U|\Psi>, \quad \tilde{H} = UH^{Ini}U^+ - i\hbar U\tfrac{d}{dt}U^+, \quad i\hbar\frac{d}{dt}|\tilde{\Psi}> = \tilde{H}|\tilde{\Psi}>, \quad U = e^{-iS}, \quad S^+ = S,$$

$$S = S^{(1)} + S^{(2)} + \dots, \quad \tilde{H} = \tilde{H}^{(0)} + \tilde{H}^{(1)} + \tilde{H}^{(2)} + \dots, \quad \tilde{H}^{(0)} = H^A + H^F, \qquad (11)$$

$$\tilde{H}^{(1)} = H^{Int} - i[S^{(1)}, \tilde{H}^{(0)}] + \hbar dS^{(1)}/dt, \quad \tilde{H}^{(2)} = -\frac{i}{2}[S^{(1)}, H^{Int}] - \frac{i}{2}[S^{(1)}, \tilde{H}^{(1)}] - i[S^{(2)}, \tilde{H}^{(0)}] + \hbar \frac{dS^{(2)}}{dt},$$

$$\dots$$

Here, the key principle of term determination in a transformed Hamiltonian $\tilde{H}$ is the presence of appropriate fast time varying factors in relevant terms (escaping due to the transition to the interaction picture) in non-diagonal matrix elements of an effective Hamiltonian. The above mentioned Hamiltonian pictures are equivalent to each other in view of the relation $\tilde{H}^{(0)} = H^A + H^F$. The unitary transformation

$$\exp(i(H^A + H^F)t/\hbar)\tilde{H}\exp(-i(H^A + H^F)t/\hbar) = \tilde{H}^{Int}(t) + H^A + H^F,$$

$$\exp(i(H^A + H^F)t/\hbar)e^{-iS}\exp(-i(H^A + H^F)t/\hbar) = e^{-iS(t)},$$

realizes the indicated equivalence.

An effective Hamiltonian (5),(6),(8)-(10) (and (11)) is diagonal in the absence of any resonance. Resonant conditions of an atom-field interaction reduce an effective Hamiltonian (5) and the Schrödinger equation (6) to the closed system of equation describing the interaction of an electromagnetic field with resonant atomic levels alone.

The peculiarities of unitary transformations (4)-(8) and (11) for the case of quantum particle interactions with classical electromagnetic fields are laid down in monograph [5]. For the quantum electromagnetic field, the method was used in the author's works [5,16,17] for different conditions other than those considered in the present paper. In these works the conception of relevant terms and the elimination of fast varying factors from them are illustrated with a great number of examples. For the first time, the analogous method was used be Van Vleck [18] and described in textbooks [19-21]. In nonlinear optics the unitary transformation of quantum states of the system under investigation has been practically applied beginning with Takatsutji's works [22]. The method of specifying unitary transformation in Refs. [18-22] differs from the above stated one. The "closure" property of an effective Hamiltonian picture has not been discussed up to now. The mathematical background for the perturbation theory for ordinary differential equations based on the transformation method similar to the above stated ones was developed in book [23].

Assume that the electromagnetic field does not have any photons. Atoms can populate either ground (lower) energy level $|E_1>$ or excited (upper) energy level $|E_2>$, thus forming an optically allowed transition $E_2 \to E_1$. The interaction with the electromagnetic field causes transitions from the excited level to the ground one only. The characteristic frequency of the electromagnetic field of such processes is determined by the frequency $\omega_{21}$ of the indicated transition $E_2 \to E_1$, $\omega_{21} = (E_2 - E_1)/\hbar$. Therefore, there arises a resonant interaction between atoms and electromagnetic field. For the resonant interaction, the operator $H^{Int}(t)$ in the interaction picture only has the following slow time varying terms which determine $\tilde{H}^{Int(1)}(t)$:

$$\tilde{H}^{Int(1)}(t) = \sum_{i,\vec{q}} \Gamma_{\vec{q}} b_{\vec{q}}^+ d_{12} e^{i(\omega_{\vec{q}} - \omega_{21})t} |E_1>^{(i)}<E_2|^{(i)} + \sum_{i,\vec{q}} \Gamma_{\vec{q}} b_{\vec{q}} d_{21} e^{-i(\omega_{\vec{q}} - \omega_{21})t} |E_2>^{(i)}<E_1|^{(i)}. \quad (12)$$

This equation allows writing down an equation for the operator $S^{(1)}(t)$ following from Eq. (9) and containing fast time varying terms with factors $e^{\pm i(\omega_{\vec{q}} + \omega_{21})t}$ eliminated from $\tilde{H}^{Int(1)}(t)$. We solve this equation making a conventional assumption that the "atom-field" interaction is switched on adiabatically, and we find the operator $S^{(1)}(t)$ in the form

$$S^{(1)}(t) = i\sum_{\vec{q}} \Gamma_{\vec{q}} b_{\vec{q}}^+ \sum_{i,kj}{}' |E_k>^{(i)}<E_j|^{(i)} \frac{e^{i(\omega_{\vec{q}} + \omega_{kj})t} d_{kj}}{\hbar(\omega_{\vec{q}} + \omega_{kj})} - i\sum_{\vec{q}} \Gamma_{\vec{q}} b_{\vec{q}} \sum_{i,kj}{}' |E_k>^{(i)}<E_j|^{(i)} \frac{e^{-i(\omega_{\vec{q}} - \omega_{kj})t} d_{kj}}{\hbar(\omega_{\vec{q}} - \omega_{kj})}. \quad (13)$$

Formula (13) contains both resonant and non-resonant atomic states. The prime sign in the sum means the absence of resonant denominators $\omega_{\vec{q}} - \omega_{21}$.

Substituting $S^{(1)}(t)$ and $\tilde{H}^{(1)}(t)$ into (10) and retaining only slow time varying terms in the commutators, we obtain $\tilde{H}^{Int(2)}$ in the form

$$\tilde{H}^{Int(2)}(t) = \sum_{\vec{q}} \Gamma_{\vec{q}} b_{\vec{q}}^+ \sum_{\vec{q}'} \Gamma_{\vec{q}'} b_{\vec{q}'} e^{-i(\omega_{\vec{q}'} - \omega_{\vec{q}})t} \sum_{i,k} \tfrac{1}{2}(\Pi_k(\omega_{\vec{q}}) + \Pi_k(\omega_{\vec{q}'}))|E_k>^{(i)}<E_k|^{(i)} +$$

$$+ \sum_{\vec{q}} \Gamma_{\vec{q}}^2 \sum_{i,kj} \frac{|d_{kj}|^2}{\hbar(\omega_{kj} - \omega_{\vec{q}})} |E_k>^{(i)}<E_k|^{(i)} - \quad (14)$$

$$- \sum_{\vec{q}} \Gamma_{\vec{q}}^2 \sum_{i \neq i',kj}{}' |E_k>^{(i)}<E_j|^{(i)} |E_j>^{(i')}<E_k|^{(i')} \frac{|d_{kj}|^2}{\hbar(\omega_{\vec{q}} - \omega_{kj})},$$

where the conventional parameters of the optical resonance theory are introduced [5]

$$\Pi_k(\omega) = \sum_{jk} \frac{|d_{kj}|^2}{\hbar} \left( \frac{1}{\omega_{kj} + \omega} + \frac{1}{\omega_{kj} - \omega} \right).$$

The rest terms in the formula (10) (after separating $\tilde{H}^{Int(2)}(t)$ in the form (14)) define the operator $S^{(2)}(t)$. However, the operator $S^{(2)}(t)$ is not required and omitted in the following.

The operator (14) is the sum of three operators

$$\tilde{H}^{Int(2)}(t) = H^{Stark}(t) + H^{Lamb} + V^{Ex},$$

$$H^{Stark}(t) = \sum_{\vec{q}} \Gamma_{\vec{q}} b_{\vec{q}}^+ \sum_{\vec{q}'} \Gamma_{\vec{q}'} b_{\vec{q}'} e^{-i(\omega_{\vec{q}'} - \omega_{\vec{q}})t} \sum_{i,k} \frac{1}{2} (\Pi_k(\omega_{\vec{q}}) + \Pi_k(\omega_{\vec{q}'})) |E_k>^{(i)}<E_k|^{(i)},$$

$$H^{Lamb} = \sum_{\vec{q}} \Gamma_{\vec{q}}^2 \sum_{i,kj} \frac{|d_{kj}|^2}{\hbar(\omega_{kj} - \omega_{\vec{q}})} |E_k>^{(i)}<E_k|^{(i)},$$

$$V^{Ex} = -\sum_{\vec{q}} \Gamma_{\vec{q}}^2 \sum_{i \neq i', kj} {}' |E_k>^{(i)}<E_j|^{(i)} |E_j>^{(i')}<E_k|^{(i')} \frac{|d_{kj}|^2}{\hbar(\omega_{\vec{q}} - \omega_{kj})}.$$

The first one $H^{Lamb}$ is regarderd as the Lamb operator. It describes the Lamb level shifts of a single atom [1]. The second operator $H^{Stark}(t)$ is referred to as the Stark interaction operator. The third operator $V^{Ex}(t)$ describes the interaction between atoms of dipole-dipole type with excitation exchange.

The Lamb operator is diagonal and can be eliminated by unitary transformation

$$|\tilde{\tilde{\Psi}}(t)> = \exp(iH^{Lamb}t/\hbar)|\tilde{\Psi}(t)>, \quad i\hbar\frac{d}{dt}|\tilde{\tilde{\Psi}}(t)> = \tilde{\tilde{H}}^{Int}(t)|\tilde{\tilde{\Psi}}(t)>,$$

$$\tilde{\tilde{H}}^{Int}(t) = \exp(iH^{Lamb}t/\hbar)\tilde{H}^{Int}(t)\exp(-iH^{Lamb}t/\hbar) -$$
$$- i\hbar\exp(iH^{Lamb}t/\hbar)\frac{d}{dt}\exp(-iH^{Lamb}t/\hbar) = \exp(iH^{Lamb}t/\hbar)\tilde{H}^{Int}(t)\exp(-iH^{Lamb}t/\hbar) - H^{Lamb},$$

$$\tilde{\tilde{H}}^{Int}(t) = \tilde{\tilde{H}}^{Int(1)}(t) + \tilde{\tilde{H}}^{Int(2)}(t).$$

We now rewrite the effective Hamiltonian in terms of resonant and non-resonant atomic levels, and present it as the sum of four operators,

$$\tilde{\tilde{H}}^{Int}(t) = \sum_{i,\vec{q}} \Gamma_{\vec{q}} b_{\vec{q}}^+ d_{12} e^{i(\omega_{\vec{q}} - \omega'_{21})t} |E_1>^{(i)}<E_2|^{(i)} + \sum_{i,\vec{q}} \Gamma_{\vec{q}} b_{\vec{q}} d_{21} e^{-i(\omega_{\vec{q}} - \omega'_{21})t} |E_2>^{(i)}<E_1|^{(i)} +$$

$$+ \sum_{\vec{q}} \Gamma_{\vec{q}} b_{\vec{q}}^+ \sum_{\vec{q}'} \Gamma_{\vec{q}'} b_{\vec{q}'} e^{-i(\omega_{\vec{q}'} - \omega_{\vec{q}})t} \sum_{i,k} \frac{1}{2}(\Pi_k(\omega_{\vec{q}}) + \Pi_k(\omega_{\vec{q}'})) |E_k>^{(i)}<E_k|^{(i)} + V(t) =$$

$$= H^{Int-TL}(t) + H^{Nonres}(t) + V^{TL-Ex} + V^{Nonres}.$$

The first one $H^{Int-TL}(t)$ describes two resonant atomic levels and transitions between them due to the interaction with the electromagnetic field,

$$H^{Int-TL}(t) = \sum_{i,\vec{q}} \Gamma_{\vec{q}} b_{\vec{q}}^+ d_{12} e^{i(\omega_{\vec{q}} - \omega'_{21})t} |E_1>^{(i)}<E_2|^{(i)} + \sum_{i,\vec{q}} \Gamma_{\vec{q}} b_{\vec{q}} d_{21} e^{-i(\omega_{\vec{q}} - \omega'_{21})t} |E_2>^{(i)}<E_1|^{(i)} +$$

$$+ \sum_{\vec{q}} \Gamma_{\vec{q}} b_{\vec{q}}^+ \sum_{\vec{q}'} \Gamma_{\vec{q}'} b_{\vec{q}'} e^{-i(\omega_{\vec{q}'} - \omega_{\vec{q}})t} \sum_{\substack{i,\\k=1,2}} \frac{1}{2}(\Pi_k(\omega_{\vec{q}}) + \Pi_k(\omega_{\vec{q}'})) |E_k>^{(i)}<E_k|^{(i)}.$$

The second operator $H^{Nonres}(t)$ characterizes the Stark interaction of non-resonant levels,

$$H^{Nonres}(t) = \sum_{\vec{q}} \Gamma_{\vec{q}} b_{\vec{q}}^+ \sum_{\vec{q}'} \Gamma_{\vec{q}'} b_{\vec{q}'} e^{-i(\omega_{\vec{q}'} - \omega_{\vec{q}})t} \sum_{\substack{i,\\k \neq 1,2}} \frac{1}{2}(\Pi_k(\omega_{\vec{q}}) + \Pi_k(\omega_{\vec{q}'})) |E_k>^{(i)}<E_k|^{(i)}$$

The third and fourth operators $V^{TL}$ and $V^{Nonres}$ represent terms of the atom-atom interaction operator $V$. The interaction between the resonant atomic states $|E_1>$ and $|E_2>$ is described by $V^{TL-Ex}$,

$$V^{TL-Ex} = -\sum_{\vec{q}} \Gamma_{\vec{q}}^2 \sum_{i \neq i'} |E_1>^{(i)}<E_2|^{(i)} |E_2>^{(i')}<E_1|^{(i')} \frac{|d_{21}|^2}{\hbar(\omega_{\vec{q}} + \omega_{21})}.$$

All the rest terms in $V$ are designated as $V^{Nonres}$.

We denoted as $\omega'_{21}$ the resonant transition frequency allowing for the Lamb shifts,

$$\omega'_{21} = \omega_{21} + \sum_{\vec{q}} \Gamma_{\vec{q}}^2 \sum_j \frac{|d_{2j}|^2}{\hbar^2(\omega_{2j} - \omega_{\vec{q}})} - \sum_{\vec{q}} \Gamma_{\vec{q}}^2 \sum_j \frac{|d_{1j}|^2}{\hbar^2(\omega_{1j} - \omega_{\vec{q}})}.$$

Furthermore, the prime at the resonant transition frequency will be omitted.

The accepted assumption related to allowed levels populations and quantum transitions permits developing the above stated theory based only on the operator $H^{Int-TL}(t) + V^{TL-Ex}$.

Thus, the dynamics of the ensemble of identical atoms in the photon-free electromagnetic field is reduced to the dynamics of the ensemble of two-level identical atoms and quantized electromagnetic field presented by the state vector $|\Psi^{TL+F}(t)>$ obeying equations

$$i\hbar \frac{d}{dt}|\Psi^{TL+F}(t)> = \{H^{Tr}(t) + H^{St}(t) + V^{TL-Ex}\}|\Psi^{TL+F}(t)>, \qquad (15)$$

$$H^{Tr}(t) = \sum_{\vec{q}} \Gamma_{\vec{q}} b_{\vec{q}}^+ d_{12} e^{i(\omega_{\vec{q}} - \omega_{21})t} R_- + \sum_{\vec{q}} \Gamma_{\vec{q}} b_{\vec{q}} d_{21} e^{-i(\omega_{\vec{q}} - \omega_{21})t} R_+,$$

$$H^{St}(t) = \sum_{\vec{q}} \Gamma_{\vec{q}} b_{\vec{q}}^+ \sum_{\vec{q}'} \Gamma_{\vec{q}'} b_{\vec{q}'} e^{-i(\omega_{\vec{q}} - \omega_{\vec{q}'})t} \{\Pi_+(\omega_{\vec{q}}, \omega_{\vec{q}'}) \frac{N_a}{2} + \Pi_-(\omega_{\vec{q}}, \omega_{\vec{q}'}) R_3\},$$

$$V^{TL-Ex} = -\sum_{\vec{q}} \frac{\Gamma_{\vec{q}}^2 |d_{21}|^2}{\hbar(\omega_{\vec{q}} + \omega_{21})} (R_- R_+ + R_+ R_- - N_a).$$

The Stark interaction parameters $\Pi_\pm(\omega, \omega')$ were involved as in

$$\Pi_\pm(\omega, \omega') = \tfrac{1}{2}\{\Pi_1(\omega) + \Pi_1(\omega') \pm (\Pi_2(\omega) + \Pi_2(\omega'))\}.$$

The operators $R_\pm$ and $R_3$ take the form:

$$R_3 = \tfrac{1}{2}\sum_i (|E_2>^{(i)}<E_2|^{(i)} - |E_1>^{(i)}<E_1|^{(i)}),\ R_- = \sum_i |E_1>^{(i)}<E_2|^{(i)},\ R_+ = \sum_i |E_2>^{(i)}<E_1|^{(i)},$$

and obey the commutation relation of the *su*(2) algebra

$$[R_3, R_\pm] = \pm R_\pm,\ [R_+, R_-] = 2R_3.$$

We will consider the initial states of two-level atoms $|\Psi_0^{TL}>$ and electromagnetic field $|\Psi_0^F>$ to be non-correlated with each other $|\Psi_0^{TL+F}> = |\Psi_0^{TL}> \otimes |\Psi_0^F>$. The field states corresponding different wave vectors are non-correlated either and are photon-free

$$<\Psi_0^F|b_{\vec{q}}^+ b_{\vec{q}'}|\Psi_0^F> = 0,\ <\Psi_0^F|b_{\vec{q}} b_{\vec{q}'}^+|\Psi_0^F> = \delta_{\vec{q}\vec{q}'}, \qquad (16)$$

$$<\Psi_0^F|b_{\vec{q}} b_{\vec{q}'}|\Psi_0^F> = <\Psi_0^F|b_{\vec{q}}^+ b_{\vec{q}'}^+|\Psi_0^F> = 0, \qquad (17)$$

Besides, $<\Psi_0^F|b_{\vec{q}}|\Psi_0^F> = <\Psi_0^F|b_{\vec{q}}^+|\Psi_0^F> = 0$. Thus, the electromagnetic field involved is a photon-free bath.

The solution to Eq.(15) is presented with the help of the evolution operator $U(t)$ ($I$ - is a unity operator):

$$|\Psi^{TL+F}(t)> = U(t)|\Psi_0^{TL+F}>,\ U(0) = I,$$

$$i\hbar \frac{d}{dt} U(t) = (H^{Tr}(t) + H^{St}(t) + V^{TL-Ex})U(t). \qquad (18)$$

Equations (15)-(18) represent the basis of analyzing collective spontaneous emission with allowance for the Stark interaction by any known method.

## 3. THE MARKOV APPROXIMATION
## AND THE RELATED QUANTUM STOCHASTIC DIFFERENTIAL EQUATION
## FOR THE SYSTEM EVOLUTION OPERATOR

Now, let us express the main equations (15)-(18) in the form suitable for further application of the QSDE method. First of all, we will write Eqs.(15)-(18) in a dimensionless form. The resonant transition frequency $\omega_{21}$ will serve as characteristic frequency, and the value of $\omega_{21}^{-1}$ will be treated as characteristic time. The value of $d_{12}$ will be considered to be real. We introduce the dimensionless time $\tau = \omega_{21} t$ and frequencies $\nu = \omega_{\vec{q}}/\omega_{21}$, $\nu' = \omega_{\vec{q}'}/\omega_{21}$. The wave vector $\vec{q}$ is presented with the help of the unity vector $\vec{n}$, $\vec{q} = \vec{n}\nu\omega_{21}/c$. We replace summation with integration

$$\sum_{\vec{q}} \to \left(\frac{\ell\omega_{21}}{2\pi c}\right)^3 \int_0^\infty 4\pi\nu^2 d\nu \int \frac{d\Omega_{\vec{n}}}{4\pi}.$$

We will denote by $\int d\Omega_{\vec{n}} \equiv \int d\Omega_{\vec{q}}$ the integration over various wave vector orientations. The following dimensionless values and operators are introduced

$$|\Psi^{TL+F}(\tau)> \equiv |\Psi^{TL+F}(\tau\omega_{21}^{-1})>, \quad b_\nu = \mu \frac{\sqrt{\ell^3}}{\pi\sqrt{2}} \left(\frac{\omega_{21}}{c}\right)^{3/2} \nu \int \frac{d\Omega_{\vec{n}}}{4\pi} b_{\vec{n}\nu\omega_{21}/c}, \quad U(\tau) \equiv U(\tau\omega_{21}^{-1}),$$

$$H^{Tr}(\tau) = \frac{1}{\sqrt{2\pi}} \int_0^\infty d\nu b_\nu^+ e^{i(\nu-1)\tau} \chi(\nu) R_- + \frac{1}{\sqrt{2\pi}} \int_0^\infty d\nu b_\nu e^{-i(\nu-1)\tau} \chi(\nu) R_+, \qquad (19)$$

$$H^{St}(\tau) = \frac{1}{2\pi} \int_0^\infty d\nu b_\nu^+ e^{i(\nu-1)\tau} \int_0^\infty d\nu' b_{\nu'} e^{-i(\nu'-1)\tau} \{\eta_+(\nu,\nu')\frac{N_a}{2} + \eta_-(\nu,\nu')R_3\}, \qquad (20)$$

$$V = -\tfrac{\kappa}{2}(R_- R_+ + R_+ R_- - N_a),$$

$$\chi(\nu) = \frac{\sqrt{2}\omega_{21} d_{12}}{\mu c^{3/2} \sqrt{\hbar}} \nu, \quad \eta_\pm(\nu,\nu') = \chi(\nu)\chi(\nu') \frac{\Pi_\pm(\omega_{21}\nu,\omega_{21}\nu')}{d_{12}^2/(\hbar\omega_{21})}, \quad \kappa = \sum_{\vec{q}} \frac{\Gamma_{\vec{q}}^2 |d_{21}|^2}{\hbar^2(\omega_{\vec{q}} + \omega_{21})\omega_{21}}.$$

The correction parameter $\mu$ was introduced for the following reason. The value $\mu = 1$ corresponds to the replacement of integration over a quantization cube to a solid sphere in the sequence of transformations

$$[b_{\vec{q}}, b_{\vec{q}'}^+] = \delta_{\vec{q}\vec{q}'} = \frac{1}{\ell^3} \int d\vec{r} e^{i(\vec{q}-\vec{q}')\vec{r}},$$

$$[\int \frac{d\Omega_{\vec{q}}}{4\pi} b_{\vec{q}}, \int \frac{d\Omega_{\vec{q}'}}{4\pi} b_{\vec{q}'}^+] = \frac{1}{\ell^3} \int \frac{d\Omega_{\vec{q}}}{4\pi} \int \frac{d\Omega_{\vec{q}'}}{4\pi} \int d\vec{r} e^{i(\vec{q}-\vec{q}')\vec{r}} = \frac{1}{\ell^3} \int 4\pi r^2 dr \int \frac{d\Omega_{\vec{q}}}{4\pi} e^{iqr\cos\theta} \int \frac{d\Omega_{\vec{q}'}}{4\pi} e^{-iq'r\cos\theta'} =$$

$$= \frac{1}{\ell^3} \int 4\pi dr \frac{1}{i2q}(e^{iqr} - e^{-iqr}) \frac{1}{i2q'}(e^{iq'r} - e^{-iq'r}) \to \frac{1}{\ell^3} \frac{2\pi^2}{qq'} \delta(q-q'), \ell \to \infty, q > 0, q' > 0.$$

Parameter $\chi$ will be seen to determine the rate of the Langevin spontaneous decay at $\eta_\pm = 0$ whereas parameters $\eta_\pm$ tend to define the non-Langevin factor of spontaneous emission suppression. In section 5 we will obtain the spontaneous decay rate for a singly excited atomic ensemble, which, for the case of a single particle, will differ from the conventional rate for the Langevin spontaneous emission [1] by a numerical factor. The value $\mu = \sqrt{3}$ corrects this divergence. For all applications of the QSDE to spontaneous emission (e.g.,[11-14]) parameter $\chi$ was treated as a phenomenological one. (However, detailed examination of this divergence is not the subject of the present paper). The above mentioned divergence does not distort the theory if we think the values $\chi$ and $\eta_\pm$ to be phenomenological parameters according to the QSDE theory of the Langevin spontaneous emission.

Now we can write down the dimensionless form of the main equations as

$$\frac{d}{d\tau}|\Psi^{TL+F}(\tau)> = -i(H^{Tr}(\tau) + H^{St}(\tau))|\Psi^{TL+F}(\tau)>, \qquad (21)$$

$$\frac{d}{d\tau}U(\tau) = -i(H^{Tr}(\tau) + H^{St}(\tau) + V)U(\tau), \quad U(0) = I, \qquad (22)$$

$$<\Psi_0^F | b_\nu^+ b_{\nu'} | \Psi_0^F > = <\Psi_0^F | b_\nu b_{\nu'} | \Psi_0^F > = <\Psi_0^F | b_\nu^+ b_{\nu'}^+ | \Psi_0^F > = 0, \quad <\Psi_0^F | b_\nu b_{\nu'}^+ | \Psi_0^F > = \delta(\nu - \nu').$$

For an ordinary atom, the Stark interaction parameters $\eta_\pm(\nu,\nu')$ are small compared with parameters $\chi(\nu)$ determining the Rabi frequency as seen from

$$\eta_{\pm}(\nu,\nu') = \chi(\nu)\chi(\nu')\frac{\Pi_{\pm}(\omega_{\bar{q}},\omega_{\bar{q}'})}{d_{21}^2/(\hbar\omega_{21})}.$$

Then, $\eta_{\pm} \ll \chi \ll 1$ as $\Pi_{\pm} \sim d_{12}^2/(\omega_{21}\hbar)$ and $\chi \ll 1$, for the expansion (8) to be reasonable. However, the relation $\Pi_{\pm} \gg d_{12}^2/(\omega_{21}\hbar)$ can be true if the anomalous smallness of the resonant transition dipole moment is possible due to any reason. As a result one can expect the values to be $\eta_{\pm} \sim 1$. Several models of two-photon transitions [6] are also characterized by $\eta_{\pm} \sim 1$. Moreover, the Stark interaction parameter $\eta_{+}$ is contained in (20) together with the number of atoms in ensemble $N_a$ as multiplicand. Therefore, the Stark interaction parameters $\eta_{\pm}$ can make an impact on atomic dynamics in the case of $N_a \gg 1$, or under some special conditions mentioned above. Therefore, the value $\eta_{+}N_a$ will be considered to be of order of unit, while $\chi \ll 1$.

Finally, we present the formal solution to Schrödinger Eq.(22) for the evolution operator $U(\tau)$ in terms of $\bar{T}$ exponent

$$U(t) = I + (-i)\int_0^\tau (H^{Tr}(\tau') + H^{St}(\tau') + V)d\tau' +$$
$$+ (-i)^2 \int_0^\tau \int_0^{\tau'} (H^{Tr}(\tau') + H^{St}(\tau') + V)(H^{Tr}(\tau'') + H^{St}(\tau'') + V)d\tau'd\tau'' + \ldots = \quad (23)$$
$$= \bar{T}\exp\left(-i\int_0^\tau (H^{Tr}(\tau') + H^{St}(\tau') + V)d\tau'\right).$$

Now we will introduce new values and make new basic assumptions characterizing the QSDE method. We will define operators

$$b(\tau) = \frac{1}{\sqrt{2\pi}}\int_{-\infty}^{\infty} d\nu e^{-i(\nu-1)\tau}b_\nu, \quad b^+(\tau) = \frac{1}{\sqrt{2\pi}}\int_{-\infty}^{\infty} d\nu e^{i(\nu-1)\tau}b_\nu^+$$

$$B(\tau) = \int_0^\tau d\tau' b(\tau'), \quad B^+(\tau) = \int_0^\tau d\tau' b^+(\tau'), \quad \Lambda(\tau) = \int_0^\tau d\tau' b^+(\tau')b(\tau') \quad (24)$$

supposing that the integration limits in $b(\tau)$ range from $-\infty$ to $+\infty$, rather than from 0 to $+\infty$. This noteworthy assumption leads to the following relations

$$[b(\tau),b^+(\tau')] = \delta(\tau-\tau'), \quad [B(\tau),B^+(\tau)] = \tau, \quad [B(\tau_1),B^+(\tau_2)] = \int_0^{\tau_1}d\tau'\int_0^{\tau_2}d\tau''\delta(\tau'-\tau'') = \min(\tau_1,\tau_2).$$

Assume that the parameters of the direct transition $\chi(\nu)$ and the Stark level shifts $\eta_{\pm}(\nu,\nu')$ are not affected by the frequency $\omega$

$$\nu = 1, \quad \chi(\nu) = const = \chi(1) \equiv \chi, \quad \eta_{\pm}(\nu,\nu') = const = \eta_{\pm}(1,1) \equiv \eta_{\pm}. \quad (25)$$

The values introduced (24) as well as assumptions (25) allow writing the interaction operators (19) and (20) in the form

$$H^{Tr}(\tau)d\tau = \chi R_{+}dB(\tau) + \chi R_{-}dB^+(\tau), \quad (26)$$

$$H^{St}(\tau)d\tau = (\eta_{+}\frac{N_a}{2} + \eta_{-}R_3)d\Lambda(\tau), \quad (27)$$

$$dB(\tau) = B(\tau+d\tau) - B(\tau), \quad dB^+(\tau) = B^+(\tau+d\tau) - B^+(\tau), \quad d\Lambda(\tau) = \Lambda(\tau+d\tau) - \Lambda(\tau). \quad (28)$$

The applied approximations are the Markov conditions, namely, the dynamics of the electromagnetic broadband field (16),(17) is determined by the field state at a point time and is not affected by the field state of the previous moments of time [1,11,13,14]. The approximations were used in all previous works [11-17] with no account of the Stark interaction. Conditions (25) along with the correction parameter $\mu = \sqrt{3}$ in the definition of $\chi$ provide the correct value of the spontaneous decay rate through the resonant atomic transition characteristics (with no allowance for photon polarization).

For the Markov conditions, Eqs. (21)-(23) have proved to be mathematically incorrect [11,13]. It is evident from the detailed consideration of integrals contained in formula (23). We will consider these integrals to be taken in the Ito form.

$$\int_0^\tau \varphi(\tau')dB^+(\tau') = \lim_{N \to \infty} \sum_{i=1}^N \varphi(\tau_{i-1})(B^+(\tau_i) - B^+(\tau_{i-1})),$$

where the limit is taken as the mean-square one [11,13,14]. Here, $0 < \tau_1 < \tau_2 < ... < \tau_{N-1} < \tau$ with $\tau_0 = 0$ and $\tau_N = \tau$, the maximum of the time intervals $\tau_i - \tau_{i-1}$ tends to zero, with the number of time interval $N$ tending to infinity. The values $\varphi(\tau)$ are nonanticipating, i.e., statistically independent of subsequent behavior of $B(\tau)$ and $B^+(\tau)$. In the mathematical way, it is expressed as

$$[\varphi(\tau), dB(\tau)] = [\varphi(\tau), dB^+(\tau)] = [\varphi(\tau), d\Lambda(\tau)] = 0.$$

The Ito quantum stochastic differential equation

$$d\varphi(\tau) = \alpha(\varphi(\tau), \tau)dB(\tau) + \beta(\varphi(\tau), t)dB^+(\tau) + \varepsilon(\varphi(\tau), \tau)d\Lambda(\tau) + \gamma(\varphi(\tau), \tau)d\tau,$$

is the equation for which the integral relation is true

$$\varphi(\tau) - \varphi(0) = \int_0^\tau \alpha(\tau')dB(\tau') + \int_0^\tau \beta(\tau')dB^+(\tau') + \int_0^\tau \varepsilon(\tau')d\Lambda(\tau') + \int_0^\tau \gamma(\tau')d\tau',$$

where the stochastic integrals are interpreted in terms of the Ito form. The differentials $d\varphi(\tau)$, $dB(\tau)$, $dB^+(\tau)$ and $d\Lambda(\tau)$ are referred to as the Ito differentials, or the Ito increments.

Hudson and Parthasarathy [10] (see, also, Refs. [24,25]) concluded that the Ito differentials (28) satisfy the below stated algebra

$$d\Lambda(\tau)d\Lambda(\tau) = d\Lambda(\tau), \ d\Lambda(\tau)dB^+(\tau) = dB^+(\tau), \ dB(\tau)d\Lambda(\tau) = dB(\tau), \ dB(\tau)dB^+(\tau) = d\tau. \quad (29)$$

$$d\Lambda(\tau)dB(\tau) = d\Lambda(\tau)d\tau = dB^+(\tau)d\Lambda(\tau) = dB^+(\tau)d\tau = dB(\tau)d\tau = 0.$$

The operators $B(\tau)$, $B^+(\tau)$ and $\Lambda(\tau)$ define the Wiener quantum process $Q(\tau)$ and the Poisson quantum process $N(\tau)$ according to [24,25]

$$Q(\tau) = B(\tau) + B^+(\tau), \ N(\tau) = \Lambda(\tau) + i(B^+(\tau) - B(\tau)).$$

The operators $dB(\tau)$, $dB^+(\tau)$ and $d\Lambda(\tau)$ are the increments of annihilation, creation and counting processes determining the Wiener quantum process $Q(\tau)$ and the Poisson quantum process $N(\tau)$. Furthermore, we will think of the Wiener quantum processes as the operators $B(\tau)$ and $B^+(\tau)$ (or $dB(\tau)$ and $dB^+(\tau)$), the Poisson quantum process as $\Lambda(\tau)$ (or $d\Lambda(\tau)$), which does not lead to any misunderstanding.

The Hudson-Parthasarathy algebra (29) allows us to provide a correct mathematical expression for the evolution operator equation. Consider the Ito differential $dU(\tau)$:

$$dU(\tau) \equiv U(\tau + d\tau) - U(\tau).$$

If Eq. (23) is taken in the form

$$U(\tau) = \lim_{N \to \infty} \exp(\frac{H^{Tr}(\tau_{N-1}) + H^{St}(\tau_{N-1}) + V}{i}(\tau_N - \tau_{N-1})) ...... \exp(\frac{H^{Tr}(\tau_0) + H^{St}(\tau_0) + V}{i}(\tau_1 - \tau_0)),$$

then

$$dU(\tau) = \{\exp(-i(\chi R_+ dB(\tau) + \chi R_- dB^+(\tau) + (\eta_+ \frac{N_a}{2} + \eta_- R_3)d\Lambda(\tau) + Vd\tau)) - 1\}U(\tau).$$

This expression shows the unitary property of the evolution operator and the Ito differentiation rule

$$d(U(\tau)U^+(\tau)) = (dU(\tau))U^+(\tau) + U(\tau)dU^+(\tau) + (dU(\tau))(dU^+(\tau)).$$

Expanding the exponent in series and applying the Hudson-Parthasarathy algebra (29) we obtain the Ito equation for the evolution operator

$$dU(\tau) = A_0 dt U(\tau) + A_+ dB(\tau) U(\tau) + A_- dB^+(\tau) U(\tau) + A_\Lambda d\Lambda(\tau) U(\tau) - iV d\tau U(\tau), \qquad (30)$$

$$dU^+(\tau) = U^+(\tau) A_0^+ d\tau + U^+(\tau) dB^+(\tau) A_+^+ + U^+(\tau) dB(\tau) A_-^+ + U^+(\tau) d\Lambda(\tau) A_\Lambda^+ + iU^+(\tau) V d\tau,$$

$$A_0 = \chi^2 R_+ \frac{e^{-i(\eta_+ \frac{N_a}{2} + \eta_- R_3)} - 1 + i(\eta_+ \frac{N_a}{2} + \eta_- R_3)}{(\eta_+ \frac{N_a}{2} + \eta_- R_3)^2} R_-, \quad A_- = \frac{e^{-i(\eta_+ \frac{N_a}{2} + \eta_- R_3)} - 1}{\eta_+ \frac{N_a}{2} + \eta_- R_3} \chi R_-, \quad A_+ = \chi R_+ \frac{e^{-i(\eta_+ \frac{N_a}{2} + \eta_- R_3)} - 1}{\eta_+ \frac{N_a}{2} + \eta_- R_3},$$

$$A_\Lambda = e^{-i(\eta_+ \frac{N_a}{2} + \eta_- R_3)} - 1.$$

The operators

$$\frac{e^{-i(\eta_+ \frac{N_a}{2} + \eta_- R_3)} - 1 + i(\eta_+ \frac{N_a}{2} + \eta_- R_3)}{(\eta_+ \frac{N_a}{2} + \eta_- R_3)^2}, \quad \frac{e^{-i(\eta_+ \frac{N_a}{2} + \eta_- R_3)} - 1}{\eta_+ \frac{N_a}{2} + \eta_- R_3}$$

are interpreted as the Taylor series of the corresponding functions of $x$: $(\eta_+ \frac{N_a}{2} + \eta_- R_3) \to x$ with subsequent reverse substitution $x \to (\eta_+ \frac{N_a}{2} + \eta_- R_3)$.

In absence of the Stark interaction $\eta_\pm = 0$ for a single quantum particle $N_a = 1$, $V = 0$, Eq.(30) coincides with the familiar case and describes the Langevin atomic relaxation [13]. The Langevin relaxation type is determined by the Langevin form of the quantum stochastic differential equation defined by the Wiener quantum processes $dB(t)$ and $dB^+(t)$ alone, in the absence of the Poisson quantum process $d\Lambda(t)$. Dependence of the evolution operator on $d\Lambda(t)$ is a sign of the non-Langevin process manifestation [24,25]. At $\eta_+ = 0$ and $N_a = 1$ Eq. (30) coincides with the non-Langevin equation derived in [6].

## 4. MASTER EQUATION FOR THE ATOMIC ENSEMBLE DENSITY MATRIX

The master equation for the density matrix of atomic ensemble and electromagnetic field $\rho(\tau) = U(\tau) | \Psi_0^{TL+F} >< \Psi_0^{TL+F} | U^+(\tau)$ is derived from the quantum stochastic differential equation for the evolution operator (30) by calculating the increment

$$d\rho(\tau) = \rho(\tau + d\tau) - \rho(\tau) =$$
$$= dU(\tau) | \Psi_0^{TL+F} >< \Psi_0^{TL+F} | U^+(\tau) + U(\tau) | \Psi_0^{TL+F} >< \Psi_0^{TL+F} | dU^+(\tau) + dU(\tau) | \Psi_0^{TL+F} >< \Psi_0^{TL+F} | dU^+(\tau).$$

The use of the Hudson-Parthasarathy algebra gives rise to

$$d\rho(\tau) = -i[V, \rho(\tau)] d\tau + A_0 d\tau \rho(\tau) + A_+ dB(\tau) \rho(\tau) + A_- dB^+(\tau) \rho(\tau) + A_\Lambda d\Lambda(\tau) \rho(\tau) +$$
$$+ \rho(\tau) A_0^+ d\tau + \rho(\tau) dB^+(\tau) A_+^+ + \rho(\tau) dB(\tau) A_-^+ + \rho(\tau) d\Lambda(\tau) A_\Lambda^+ +$$
$$+ A_+ dB(\tau) \rho(\tau) dB^+(\tau) A_+^+ + A_+ dB(\tau) \rho(\tau) dB(\tau) A_-^+ + A_+ dB(\tau) \rho(\tau) d\Lambda(\tau) A_\Lambda^+ +$$
$$+ A_- dB^+(\tau) \rho(\tau) dB^+(\tau) A_+^+ + A_- dB^+(\tau) \rho(t) dB(\tau) A_-^+ + A_- dB^+(\tau) \rho(\tau) d\Lambda(\tau) A_\Lambda^+ +$$
$$+ A_\Lambda d\Lambda(\tau) \rho(\tau) dB^+(\tau) A_+^+ + A_\Lambda d\Lambda(\tau) \rho(\tau) dB(\tau) A_-^+ + A_\Lambda d\Lambda(\tau) \rho(\tau) d\Lambda(\tau) A_\Lambda^+.$$

The master equation for the atomic ensemble only $\rho^{TL}(\tau) = Tr_F \rho(\tau)$ on account of the relations

$$Tr_F (\rho(\tau) dB(\tau)) = Tr_F (\rho(\tau) dB^+(\tau)) = Tr_F (\rho(\tau) d\Lambda(\tau)) = 0.$$

is derived in the form

$$\frac{d\rho^{TL}}{d\tau} = -i[V, \rho^{TL}] + \chi^2 a_-^{NL}(\eta_+, \eta_+, R_3) R_- \rho^{TL} R_+ a_+^{NL}(\eta_+, \eta_+, R_3) - $$
$$-\frac{\chi^2}{2}\{R_+\{a_0^{NL}(\eta_+, \eta_+, R_3) - ia_s^{NL}(\eta_+, \eta_+, R_3)\}R_- \rho^{TL} + \rho^{TL} R_+\{a_0^{NL}(\eta_+, \eta_+, R_3) + ia_s^{NL}(\eta_+, \eta_+, R_3)\}R_-\}. \tag{31}$$

This equation describes the non-Langevin collective spontaneous decay of ensemble of identical two-level atoms in a photon-free vacuum field. The non-Langevin operators are involved

$$a_0^{NL}(\eta_+, \eta_+, R_3) = 2\frac{1 - \cos(\eta_+ \frac{N_a}{2} + \eta_- R_3)}{(\eta_+ \frac{N_a}{2} + \eta_- R_3)^2}, \quad a_s^{NL}(\eta_+, \eta_+, R_3) = 2\frac{\eta_+ \frac{N_a}{2} + \eta_- R_3 - \sin(\eta_+ \frac{N_a}{2} + \eta_- R_3)}{(\eta_+ \frac{N_a}{2} + \eta_- R_3)^2},$$

$$a_\pm^{NL}(\eta_+, \eta_+, R_3) = \frac{\cos(\eta_+ \frac{N_a}{2} + \eta_- R_3) - 1}{\eta_+ \frac{N_a}{2} + \eta_- R_3} \pm i\frac{\sin(\eta_+ \frac{N_a}{2} + \eta_- R_3)}{\eta_+ \frac{N_a}{2} + \eta_- R_3}.$$

In the absence of the Stark interaction $\eta_\pm = 0$, the non-Langevin operators are proportional to the unity operator $I$

$$a_0^{NL}(\eta_+, \eta_+, R_3) = I, \quad a_s^{NL}(\eta_+, \eta_+, R_3) = 0, \quad a_\pm^{NL}(\eta_+, \eta_+, R_3) = \pm iI,$$

and the master equation (31) agrees with the familiar master equation describing collective Langevin atomic decay [13].

The non-Langevin operators obey the relation

$$a_+^{NL}(\eta_+, \eta_+, R_3) a_-^{NL}(\eta_+, \eta_+, R_3) = a_0^{NL}(\eta_+, \eta_+, R_3).$$

One can introduce the Lindblad operators

$$L_- = a_-^{NL}(\eta_+, \eta_+, R_3) R_-, \quad L_+ = R_+ a_+^{NL}(\eta_+, \eta_+, R_3), \quad L_+ L_- = R_+ a_0^{NL}(\eta_+, \eta_+, R_3) R_-,$$

and give the master equation the Lindblad form [26]

$$\frac{d\rho^{TL}}{d\tau} = \chi^2 L_- \rho^{TL} L_+ - \frac{\chi^2}{2}(L_+ L_- \rho^{TL} + \rho^{TL} L_+ L_-) - i[(H^{NL-ST} + V), \rho^{TL}], \tag{32}$$

where

$$H^{NL-ST} = -\frac{\chi^2}{2} R_+ a_s^{NL}(\eta_+, \eta_+, R_3) R_-.$$

The operator $H^{NL-ST}$ defines the atomic level shifts caused by the Stark interaction. These same shifts do differ both from the Lamb shifts included into the frequency $\omega_{21}$ and the Stark shifts represented as $<\Psi_0^F | H^{Stark}(t) | \Psi_0^F>$, and equal to zero in a photon-free electromagnetic field. For a single atom, this shift was described in [6].

Note that the operator $V$ also leads to level shifts due to the excitations exchange in atom-atom interactions of dipole-dipole type. Therefore, the operator $V$ can be regarded as an excitation-exchange operator.

The next two sections will provide solutions to Eq.(32) for the cases of singly- and fully excited atomic ensembles.

## 5. NON-LANGEVIN DECAY OF SINGLY EXCITED ATOMIC ENSEMBLE

Collective spontaneous emission of identical atoms in the common vacuum field is not easy to be completely investigated. Even a two-atom system with originally factorized different quantum states appears to be entangled as a result of spontaneous decay [27-30]. The main reason for the system behavior is due to the existence of the decoherence-free subspace [31,32]. Below we will consider simpler cases of collective spontaneous emission with the initial Dicke state[3]. For a singly excited atomic ensemble, these states will reduce to W-states [33], which are important in quantum information processing.

Let the initial atomic ensemble state at $\tau = 0$ be the following

$$|E_0^{TL}> = \frac{1}{\sqrt{N_a}} \{|E_2>^{(1)}|E_1>^{(2)} \ldots |E_1>^{(N_a)} + |E_1>^{(1)}|E_2>^{(2)} \ldots |E_1>^{(N_a)} +$$
$$= \ldots + |E_1>^{(1)}|E_1>^{(2)} \ldots |E_2>^{(N_a)}\}.$$

It is quite essential to make clear how the number of atoms $N_a$ affects the non-Langevin decay rate. Let $N_a = 2r$. Because of the symmetry properties of $|E_0^{TL}>$ with respect to permutations of atoms, $|E_0^{TL}>$ can be expressed by $|r,-r+1>$ of $N_a$-dimensional space of irreducible representation of su(2) algebra,

$$R_+|r,m-1> = \sqrt{(r+m)(r-m+1)}\,|r,m>, \quad R_-|r,m> = \sqrt{(r+m)(r-m+1)}\,|r,m-1>,$$
$$R_3|r,m> = m|r,m>, \quad -r \leq m \leq r,$$

with the Casimir operator $R^2 = R_+R_- + R_3^2 - R_3 = R_-R_+ + R_3^2 + R_3$: $R^2|r,m> = r(r+1)|r,m>$. The excitation exchange operator $V$ has the simplest form, $V = -\kappa(r^2 - R_3^2)$, for the irreducible representation space.

The equation for the density matrix $\rho_{-r+1,-r+1}^{TL}$, describing the decay of singly excited atomic ensemble, is derived from (32),

$$\frac{d\rho_{-r+1,-r+1}^{TL}}{dt} = -4\chi^2 r \frac{1-\cos(r\eta_+ - r\eta_-)}{(r\eta_+ - r\eta_-)^2} \rho_{-r+1,-r+1}^{TL}. \tag{33}$$

The singly excited atomic ensemble decays exponentially so that the population of excited state is given by

$$\rho_{-r+1,-r+1}^{TL}(\tau) = \exp\{-4\chi^2 r \frac{1-\cos(r\eta_+ - r\eta_-)}{(r\eta_+ - r\eta_-)^2} \tau\}.$$

For one atom $r = 1/2$ and $\eta = (\eta_+ - \eta_-)/2$ Eq. (32) coincides with master equation obtained in work [6], although non-Langeven spontaneous emission was investigated in [6] on the basis of a simpler model.

If the Stark interaction is absent or negligibly small $\eta_\pm = 0$, Eq. (32) describes the Langevin decay of the singly excited atomic ensemble

$$\rho_{-r+1,-r+1}^{TL}(\tau) = \exp\{-2\chi^2 r\tau\}.$$

Thus, the constant $2r\chi^2$ may be regarded as the Langevin decay rate of the singly excited atomic ensemble. It is directly proportional to the number of ensemble atoms.

The Stark interaction of atoms with vacuum photon-free electromagnetic field produces the decrease of spontaneous emission rate of atomic ensemble (in comparison with the Langevin case) at any intensity of the Stark interaction. That is the consequence of quantum interference of both spontaneous transition from the excited level to ground one, with one photon being emitted, and virtual transitions with returning to the excited level with no photon emission. As the number of ensemble atoms increases, the total intensity of the Stark interaction increases too. Despite the fact that Langevin spontaneous emission rate constantly increases as the number of atoms rises $N_a = 2r$, there exists the critical value of the number of atoms $N_a^{cr} = 4\pi/(\eta_+ - \eta_-)$ or parameter $r$

$$r^{cr} = 2\pi/(\eta_+ - \eta_-), \tag{34}$$

at which the spontaneous decay is completely suppressed and

$$\rho_{-r+1,-r+1}^{TL}(\tau) = \rho_{-r+1,-r+1}^{TL}(0) = const.$$

Let us note that the dimensionless combination of parameters $\eta_+ - \eta_-$ is determined by the value $\Pi_2(\omega)$, characterizing the Stark shift of the excited level, so

$$N_a^{cr} = \frac{2\pi d_{12}^2}{\chi^2 |\Pi_2|\omega_{21}\hbar}.$$

Work [6] was devoted to analyzing phenomena where the Stark interaction growth of a single quantum particle could be expected to get substantial. With spontaneous emission being the basic reason for decoherence producing mechanism and hindering the quantum operations, search for situations where the Stark interaction can not be neglected is of interest for quantum information processing. However, it is not a subject for discussion in the present paper.

It is important to emphasize that the quantum Poisson process $\Lambda(\tau)$ describing the Stark interaction acts as if it were an original "interaction accumulator" according to the relation $d\Lambda(\tau)d\Lambda(\tau) = d\Lambda(\tau)$. Instead of making quantum transition, with a photon emitting, a particle is involved in perpetual virtual transitions and returns to an excited level with no photon emission. It gives rise to stabilization of an excited state. To calculate such an effect in a different way [7-9] other than quantum SDE method, it is necessary to summarize an infinite series where only the first terms have been accounted previously. The same "interaction accumulator" effect has revealed in the photon counting while the radiating particle is being continuously measured [12]. The summation of the mentioned above infinite series is automatically carried out in the QSDE method due to the Hudson-Partasarathy algebra (29). It is the basic advantage of the QSDE method.

## 6. NON-LANGEVIN DECAY OF FULLY EXCITED ATOMIC ENSEMBLE

One more relatively simple case of spontaneous decay of excited atomic ensemble with the Stark interaction manifestation is the spontaneous decay of fully excited two-level particles. At $\tau = 0$ the initial state of such an atomic ensemble is

$$|E_0^{TL}> = |E_2>^{(1)}|E_2>^{(2)}...|E_2>^{(N_a)}.$$

The relaxation dynamics of this system is also given by Eq.(32) in case of state space spanned by vectors $|r,m>$, $N_a = 2r$ with the initial state $|E_0^{TL}> = |r,r>$. For the diagonal matrix elements, it is easy to obtain the equation from (32)

$$\frac{d\rho_{mm}^{TL}}{dt} = -2\chi^2 g_{m\,m-1}\frac{1-\cos(r\eta_+ + \eta_-(m-1))}{(r\eta_+ + \eta_-(m-1))^2}\rho_{mm}^{TL} + 2\chi^2 g_{m+1\,m}\frac{1-\cos(r\eta_+ + \eta_-m)}{(r\eta_+ + \eta_-m)^2}\rho_{m+1\,m+1}^{TL}, \quad (35)$$

where

$$g_{m\,m-1} = <m|R_+|m-1><m-1|R_-|m> = (r+m)(r-m+1).$$

The decay of excited atomic ensemble is of non-Langevin type if the number of ensemble atoms is sufficiently high. Here, we can also think of the existence of the critical value of the number of atoms in the ensemble

$$r^{cr} = \frac{2\pi + \eta_-}{\eta_+ + \eta_-} \cong \frac{2\pi}{\eta_+ + \eta_-}, \quad (36)$$

where $\rho_{rr}^{TL}(\tau) = const = 1$ and the atomic ensemble of fully excited two-level atoms is not decayed as a result of the excitation stabilization caused by the Stark interaction. It is a curious notion that the critical value (36) is different from the critical value (34) for the singly excited atomic ensemble and is defined by the Stark interaction parameter characterizing the ground atomic level alone. It can be seen from the corresponding expression given for the critical number of atoms

$$N_a^{cr} = \frac{2\pi d_{12}^2}{\chi^2|\Pi_1|\omega_{21}\hbar}.$$

Lastly, there are critical values $m^{cr} \cong (2\pi - r\eta_+)/\eta_-$ of atomic ensemble excitations when the collective spontaneous emission of the fully excited atomic ensemble brings to a halt and the ensemble is stabilized in the excited state with $m^{cr} + r$ excited atoms, $0 \leq m^{cr} + r \leq N_a$.

Depending of the meaning of the Stark interaction parameters $\eta_\pm$ all the above indicated critical values "break" the atomic ensemble dynamics into predetermined domains where the non-Langevin factors exert various impacts on the spontaneous emission rates, which can either increase or decrease in the course of photon emission.

Now we will consider the average intensity of collective spontaneous emission $\bar{I}(t)$ which is directly proportional to energy losses of the atomic ensemble

$$\bar{I}(t) = -\alpha \frac{d}{dt} Tr(\hbar\omega_{21} R_3 \rho^{TL}),$$

where the geometrical factor $\alpha$ is introduced and level shifts are neglected. Then

$$\bar{I}(t) = \alpha \sum_{m=-r}^{r} \hbar\omega_{21} 2\chi^2 G_{m\,m-1} \rho_{mm}^N \equiv \alpha' \sum_{m=-r}^{r} G_{m\,m-1} \rho_{mm}^N, \quad G_{m\,m-1} = g_{m\,m-1} \frac{1-\cos(r\eta_+ + \eta_-(m-1))}{(r\eta_+ + \eta_-(m-1))^2}.$$

In the simplest case $\eta_- = 0$ Eq. (35) coincides with the basic equations of the conventional superradiance theory [4], which allows us to apply the well-known analytical results to a large number of excited atoms $r \gg 1$:

$$\bar{I}(t) \approx \bar{\gamma} r^2 \text{sech}^2[\bar{\gamma} r(t-t_D)], \quad \bar{\gamma} = 2\chi^2 \alpha \hbar\omega_{21} \frac{1-\cos(\eta_+ r)}{(\eta_+ r)^2}, \quad t_D = (2\bar{\gamma}r)^{-1} \ln 2\bar{\gamma}r. \tag{37}$$

Here again one can introduce the critical value of the atom number $N_a^{cr}$ (or parameter $r^{cr} = N_a^{cr}/2$), when the decay is fully suppressed

$$r^{cr} = 2\pi/\eta_+. \tag{38}$$

At different values of the number of ensemble atoms, the collective spontaneous emission for $\eta_- = 0$ is similar to the conventional (Langevin) superradiance [4]. For the conventional superradiance, however, the larger is the number of excited atoms, the shorter is the pulse duration and time delay of superradiance. At the same time for the non-Langevin superradiance described by formula (37) one can observe the increase of superradiance duration and time delay as the number of ensemble atoms rises due to the non-Langevin factor $(1-\cos(\eta_+ r))/(\eta_+ r)^2$ renormalizing the rate $\bar{\gamma}$ of the collective spontaneous emission and establishing the main difference of the collective spontaneous emission from the conventional case under the strong Stark interaction.

The derived regularities of collective spontaneous emission of identical atoms ensemble can serve as a starting point for further detailed investigation with allowance for different atomic excitation distributions, different atomic positions, etc. In the present paper our aim was to attract the reader's attention to the phenomenon of the enhancement of the Stark interaction with vacuum while increasing the particle number in the atomic ensemble. As a result, the Stark interaction, in turn, gives rise to radical changes in the collective spontaneous emission invoking its suppression and stabilization of the excited atomic ensemble.

## 7. CONCLUSION

The main difference of the present investigation from previous works devoted to spontaneous emission is the allowance for the Stark interaction with a vacuum electromagnetic field. In the absence of such interaction, spontaneous emission is of Langevin type and is defined by the quantum Wiener process. Then, the higher is the interaction intensity with a field and/or the particle number in the atomic ensemble, the higher is the spontaneous emission rate. The Stark interaction is described by the quantum Poisson process, which determines the non-Langevin type of spontaneous emission. The role played by the Stark interaction increases as the number of atoms taking part in the collective spontaneous emission rises. The enhancement of the Stark interaction intensity and particle number in the atomic ensemble gives rise to non-linear decrease of spontaneous emission rate as compared to the Langevin decay. It has been found that there is a set of critical values for the number of ensemble atoms taking part in collective spontaneous emission when the spontaneous emission is fully suppressed and the atomic ensemble is stabilized in an excited state.

One more distinctive feature of the present work is the formulation of the effective Hamiltonian picture along with the QSDE for the evolution operator. This approach allows deriving the atomic master equation in a straightforward and elegant manner owing to both effective

Hamiltonian picture and the Hudson-Partasarathy algebra (29) for the increment of the quantum stochastic processes.

The research done in the field of spontaneous emission has provided an example of analysis of an open system in the Markov approximation in terms of combining the effective Hamiltonian of the open system and QSDE for the evolution operator. The QDSEs are formulated in the framework of the factorization of the initial state of the open system and its environment, independence of different environment modes (Eqs. (16),(17)) and the homogeneous interaction (15) combined with the definitions of quantum stochastic processes in the Markov approximation (Eq.(25)). All these requirements laid down above are related not to the initial Hamiltonian and state vector but to the transformed, or effective, Hamiltonian and transformed initial state vector (Eqs. (5) and (4)). As a result, the master equation in the Lindblad form has been derived for the atomic ensemble. The Lindblad operator has been shown to consist of the non-Langevin operators such as $a_\pm^{NL}$. The master equation contains no parameter $\Delta = \omega_\Gamma - \omega_{21}$ characterizing the feasible detuning of the central frequency $\omega_\Gamma$ of broadband quantized electromagnetic field from the resonant transition frequency $\omega_{21}$ in view of singular conditions (16) and (17), and uncertainty equation for the evolution operator. Involvement of such parameter and analysis of the case when it becomes large enough, e.g. compared with the spontaneous emission rate are impossible due to the basic requirement for the effective Hamiltonian, i.e. the lack of fast varying terms in the interaction picture. The allowance for the parameter similar to $\Delta$ means that at transition to the transformed Hamiltonian, not all fast varying terms in the interaction picture are excluded, so this Hamiltonian can not be regarded as the effective Hamiltonian. Therefore, in such circumstances one should transform the Hamiltonian so that all fast varying terms including the ones related to the parameter $\Delta$ are to be excluded. Only such Hamiltonian can be thought of as the effective Hamiltonian. This requirement differs from those of previous works [18-22] describing unitary transformation for the Hamiltonian simplification. For the first time attention has been attracted to the stated principle obedience in the work [17] where the dispersive limit for the master equation for atoms in the low Q-cavity (with losses) was considered.

The dispersive limit means that the detuning $\Delta = \omega_c - \omega_{21}$ from the resonance of the cavity mode frequency $\omega_c$ is great in comparison with the detuning-free quantum transition rate. The effective Hamiltonians for small and large values of the parameter $\Delta$ have proved to be different from each other and not correlated to each other by the limiting transition as $\Delta$ grows. Each master equation being obtained from its corresponding effective Hamiltonian is also different to each other and not related to each other by the limiting transition. Thus, the above stated principle of the absence of fast varying factors in the relevant terms defines the unique effective Hamiltonian for each physical condition as in the example of the strong resonant interaction of the cavity mode with atoms or its dispersive limit. Each effective Hamiltonian in the Markov approximation determines its own QSDE and master equation. It is a distinctive feature of a new approach towards the investigation of open systems, the notions of which were developed in works [5,16,17,27]. The present paper provides an essential development to this approach. Now the quantum Poisson process and the second order terms over the coupling with the environment have been introduced into the effective Hamiltonian and its QSDE. The approach can be successfully applied to formulation and solution to new problems in the field of nonlinear and quantum optics and open system theory, in particular allowing for various two-quantum radiating processes.

In addition, the above stated principle of the absence of fast varying factors in the relevant terms should be taken into account in the process of investigating the dynamics of open systems started from the Lindblad-type master equation. The Lindblad-type master equation is the general form of dissipative dynamics controlled by the continuous quantum dynamic semigroup [26]. However, there is no apparent restriction to the fast varying terms in the Lindblad-type master equation. Assuming the existence of fast and slow subsystems in the open system being described by the Lindblad-type master equation, it is necessary to obtain new Lindblad operators by the above stated approach.

The new effects of the spontaneous emission suppression presented in the current paper may appear to be useful in quantum information processing, providing the decoherence-free excited atomic states.

The author thanks Prof. Alexander Chebotarev for his helpful assistance in discussing mathematical questions connected with the Poisson process.